\documentclass[aps,preprint,groupedaddress,showpacs]{revtex4}

\usepackage{graphicx}
\usepackage{epsfig}

\begin{document}
\bibliographystyle{apsrev}


\title{Quantum Brownian Motion With Large Friction}

\author{Joachim~Ankerhold$^{1,2}$, Hermann~Grabert$^1$ and
 Philip~Pechukas$^{3}$}

\affiliation{${}^1$ Physikalisches Institut,
Albert-Ludwigs-Universit\"at,
 79104 Freiburg, Germany\\
${}^2$ Service de Physique de l'Etat Condense, Centre d'Etudes
de Saclay, 91191 Gif-sur-Yvette, France\\
${}^3$ Department of Chemistry, Columbia University, New York, New
York 10027, USA}

\date{\today}

\begin{abstract}
Quantum Brownian motion in the strong friction limit is
studied based on the exact path integral formulation of
dissipative systems. In this limit the time-nonlocal
reduced dynamics can be cast into an effective equation of
motion, the quantum Smoluchowski equation. For strongly
condensed phase environments it plays a similar role as
master equations in the weak coupling range. Applications
for chemical, mesoscopic, and soft matter systems are
discussed and reveal the substantial role of quantum
fluctuations.

\end{abstract}

\maketitle

{\bf Quantum Brownian motion is much more involved than its
classical analog since in general tractable equations of
motion do not exist. Progress has been made in the weak
friction regime leading to a variety of master equations.
Here, we analyze the opposite domain of very strong
friction and reveal that a generalization of the classical
Smoluchowski equation, the quantum Smoluchowski equation,
can be derived from first principles. This opens the door
to study the dynamics of strongly condensed phase systems
at lower temperatures. In particular, quantum fluctuations
turn out to play a substantial role as shown explicitly for
three examples from chemical reactions, mesoscopic
physics, and charge transfer in molecules.}

\section{Introduction}
Brownian motion, that is the fate of a heavy particle immersed in
a fluid of lighter particles, is {\it the} prototype of a
dissipative system coupled to a thermal bath with infinitely many
degrees of freedom. The work by Einstein in 1905 \cite{einstein05}
has developed a mathematical language to describe the random
motion of the particle and has uncovered the fundamental relation
between friction, diffusion and the temperature $T$ of the bath.
Half a century later, this seed had grown into the theory of
irreversible thermodynamics \cite{irr}, which governs the
relaxation and fluctuations of classical systems near
equilibrium. At that time, a new challenge had emerged, the
quantum mechanical description of dissipative systems.

In contrast to classical Brownian motion, where right from
the beginning the work by Einstein and Smoluchowski
\cite{einstein05,smoluchowski00} had allowed to consider
both weak and strong friction, for a long time, the quantum
mechanical theory could handle the limit of weak
dissipation only. In this case the interaction between the
``particle" and the ``bath" can be treated perturbatively
and one can derive a master equation for the reduced
density matrix of the ``particle" \cite{masterequation}.
This approach, based on the so--called Born and Markov
approximations, has been very successful in quite a number
of fields emerging in the fifties and sixties of the last
century, such as nuclear magnetic resonance
\cite{wangsness,redfield} and quantum optics
\cite{qmoptics}.

Roughly, a dissipative quantum system can be characterized by three
typical energy scales, an excitation energy $\hbar\omega_0$, where
$\omega_0$ is a characteristic frequency of the system, a coupling energy
$\hbar\gamma$ to the bath, where $\gamma$ is a typical damping constant,
and the thermal energy $k_BT$. The weak coupling master equation is
limited to the region $\hbar \gamma \ll \hbar\omega_0,k_BT$. This is the
case whenever the typical linewidth caused by environmental interactions
is small compared to the line separation and the thermal ``Matsubara"
frequency $2\pi k_BT/\hbar$. On the other hand, the approach will fail
for strong damping and/or low temperatures.

The work by Feynman and Vernon \cite{feynman} has shown how to take
advantage of the path integral representation of quantum mechanics to
derive an expression for the reduced density matrix of the ``particle" as
a sum over forward and backward paths, which is valid for arbitrary
damping strength and temperature. These two sets of paths arise from the
two time evolution operators in the formal expression for the
time--dependent density matrix $W(t) = \exp[-(i/\hbar)Ht] \,W(0)\,
\exp[+(i/\hbar)Ht]$, where $H$ is the Hamiltonian of the system. In the
path integral representation of the reduced density matrix the influence
of the bath emerges as time--nonlocal terms in the action governing the
path probability. These terms include also a coupling between the forward
and backward paths and make it difficult to evaluate the path integral
explicitly, although progress can be made in some cases
\cite{calde83a,grabe88}.

Everything becomes simpler in the limits of weak and strong
damping. While the former case leads to the master equation
approach mentioned above, the latter case is usually referred to
as the Smoluchowski limit. In the classical case this limit is
well understood and simplifies matter considerably, since the
momentum $P$ of a heavily damped particle with position $Q$ is a
slow variable that can be eliminated adiabatically. It is
intuitively clear, that a quantum mechanical theory cannot
entirely dispose of the variable $P$ conjugate to $Q$. In fact,
for a classical system with characteristic frequency $\omega_0$
the roots $\lambda_{\pm} = \pm i\omega_0$ of the characteristic
equation in the undamped case are in the presence of damping
turned into
\[
\lambda_{\pm}=-{\gamma\over 2}\pm\sqrt{{\gamma^2\over
4}-\omega_0^2}\,.
\]
For large friction, $\gamma\gg\omega_0$, this can be approximated
by
\[
  \lambda_{\pm}=\left\{ {-{\omega_0^2\over\gamma}
  \atop{-\gamma+{\omega_0^2\over\gamma}}}\right.
  \,.
\]
While in the strong friction limit the slow classical dynamics is
governed by the small root $\omega_0^2/\gamma$, quantum mechanics
accompanies reduced fluctuation of $Q$ with enlarged fluctuations
of the conjugate variable $P$. Both roots $\lambda_{\pm}$ are
essential to determine the fluctuation spectrum. That is why
quantum effects are not restricted to the temperature range $k_BT
\le \hbar{\omega_0^2/\gamma}$, rather, in quantities sensitive to
fluctuations, quantum effect can manifest themselves in the
strong friction limit even up to higher temperatures than in the
weak damping limit. This makes the quantum Smoluchowski limit
addressed in this article nontrivial.

The paper is organized as follows. In the next section
\ref{qse} we shortly review the formulation of dissipation
within the path integral formalism and sketch the
derivation of the Quantum Smoluchowski Equation (QSE)
\cite{ankerhold01}. The second part
(Sec.~\ref{applications}) is devoted to three specific
applications.

\section{The Quantum Smoluchowksi Equation}\label{qse}

\subsection{Reduced dynamics}

As already pointed out in the Introduction the inclusion
of quantum dissipation in a non-perturbative way has been
established only since the early 80s. The standard
approach \cite{weiss} starts from a system+reservoir model
\begin{equation}
H=H_S+H_R+H_I\label{sm1}
\end{equation}
with a system part $H_S$, an environmental part $H_R$, and
a system-bath interaction $H_I$. The reservoir (heat bath)
is mimicked by a quasi-continuum of harmonic oscillators
bilinearly coupled to the system. Dissipation appears when
one considers the reduced dynamics by properly eliminating
the bath degrees of freedom. In fact, one this way regains
in the classical limit a generalized Langevin equation. It
is thus important to realize that the only restriction
associated with the oscillator bath is the Gaussian
stochastic nature of the environment, which in turn means
that the relevant impact of the bath onto the system
dynamics is completely determined by the average and the
autocorrelation function of the bath force.

The quantum dynamics of the reduced system follows from
\begin{equation}
\rho(t)={\rm Tr}_R\left\{\exp(-i H t/\hbar)\ W(0)\ \exp(i H
t/\hbar)\right\}\label{sm2}
\end{equation}
where $W(0)$ describes the initial state of the total
system. In the ordinary Feynman-Vernon theory
\cite{feynman} this state is assumed to be a factorizing
state, $W(0)=\rho_S(0)\, \exp(-\beta H_R)/Z_R$ ($Z_R$ is
the bath partition function and $\beta=1/k_{\rm B} T$), so
that each one, system and equilibrated bath, lives in
splendid isolation at $t=0$. While this assumption may be
justified in the weak damping/high temperature limit,  it
certainly fails for moderate to strong friction and/or
lower temperatures. It can be shown explicitly that in the
classical limit even the Langevin equation is not
regained, but differs by initial boundary terms that may
persist up to long times. In the strong damping realm
considered here, the initial state must be correlated
\cite{grabe88}, i.e.\ $W(0)=\sum_j O_j \exp(-\beta H)
O_j'/Z$ with preparation operators $O_j$ and $O_j'$ acting
onto the system degree of freedom only and the total
partition function $Z$. For transparency, in the sequel we
assume these operators to be diagonal in position so that
the initial density for a one-dimensional system  reads in
position representation
\begin{equation}
\rho(q_i,q_i',t=0)=\rho_\beta(q_i,q_i')\, \lambda(q_i,q_i')\label{sm3}
\end{equation}
with the reduced thermal equilibrium density matrix $\rho_\beta(q,q')={\rm
Tr}_R\langle q|\exp(-\beta H)|q'\rangle$
  and the coordinate representation $\lambda(q,q')=\sum_j
  \langle q| O_j|q\rangle
  \langle q'| O_j'|q'\rangle$ of the
  preparation operators. For a specific initial state, one does not need to know
  the specific form of these operators (which may be very
  complicated), but rather chooses the preparation function
  properly.

The reduced quantum dynamics (\ref{sm2}) starting with an initial state
(\ref{sm3}) is now obtained within the position representation by
employing the path integral formalism. Since the bath contains harmonic
degrees of freedom only, it can be integrated out exactly and one arrives
at
\begin{equation}
\rho(q_f,q_f',t)=\int d q_i dq_i'\ J(q_f,q_f',t,q_i,q_i')\,
\lambda(q_i,q_i')
\label{sm4}
\end{equation}
where the propagating function $J(\cdot)$ is a threefold
path integral---two in real time, one in imaginary
time---over the system degree of freedom only
\begin{equation}
J(q_f,q_f',t,q_i,q_i')=\frac{1}{Z} \int {\cal D}[q] {\cal D}[q']{\cal
  D}[\bar{q}]\ {\rm e}^{i \Sigma[q,q',\bar{q}]/\hbar}\, \label{sm5}
\end{equation}
with $Z={\rm Tr}\{\exp(-\beta H)\}/Z_R$. The two real time
paths $q(s)$ and $q'(s)$ connect in time $t$ the initial
points $q_i$ and $q_i'$ with fixed end points $q_f$ and
$q_f'$, while the imaginary time path $\bar{q}(\sigma)$
runs from $q_i$ to $q_i'$ in the interval $\hbar\beta$.
The contribution of each path is weighted with an
effective action $\Sigma[q,q',\bar{q}]=S_S[q]-S_S[q']+i
\bar{S}[\bar{q}]+i \phi[q,q,',\bar{q}]$ which consists of
the actions of the bare system in real and imaginary time,
respectively, and an additional interaction contribution
(influence functional) non-local in time. The latter one
can be written as
\begin{equation}
\phi[\tilde{q}]=\int dz \int_{z>z'} dz'\ \tilde{q}(z)\, K(z-z')\,
\tilde{q}(z')+\frac{i}{2} \mu\int dz \tilde{q}(z)^2\label{sm6}
\end{equation}
where the ordered time integration is understood along the
contour:  $z=s$ from $t \to 0$,  $z=-i\tau$ from $0\to
\hbar\beta$, $z=-i\hbar\beta +s$ from $0\to t$ with
\begin{equation}
\tilde{q}(z)=\left\{
\begin{array}{lll}
q'(s)\ & \ \mbox{for}\ z=s &\ 0\leq s\leq t\\
\bar{q}(\tau)\ & \ \mbox{for} \ z=-i\tau &\ 0\leq \tau\leq \hbar\beta\\
q(s)\ &\ \mbox{for}\ z=-i\hbar\beta +s &\ 0\leq s\leq t
\end{array}
\right.\, .
\label{sm7}
\end{equation}
The effective impact of the bath is completely controlled by the
damping kernel
\begin{equation}
K(z)=\int_0^\infty \frac{d\omega}{\pi}\ I(\omega)\, \frac{{\rm
    cosh}[\omega(\hbar\beta-i z)]}{{\rm
    sinh}(\omega\hbar\beta/2)}\label{sm8}
\end{equation}
where $I(\omega)$ denotes the spectral density of the environment. As
expected $\hbar K(z)$ coincides with the autocorrelation function of the
bath force. In particular, for real times the kernel $K(s)=K'(s)+i
K''(s)$ is related to the macroscopic damping kernel entering the
classical generalized Langevin equation
\begin{equation}
\gamma(s)=\frac{2}{M} \int_0^\infty \frac{d\omega}{\pi} \
\frac{I(\omega)}{\omega}\, \cos(\omega s)\label{sm9}
\end{equation}
via $K''(s)=(M/2) d\gamma(s)/ds$ and $K'(s)\to
M\gamma(s)/\hbar\beta$ in the classical limit ($M$ is the
mass of the Brownian particle). The term with
$\mu=\lim_{\hbar\beta\to 0} \hbar\beta K(0)$ in (\ref{sm6})
gives a potential renormalization due to shifts of the
minima of the bath oscillators by the coupling to the
system.

\subsection{Quantum Smoluchowski range and thermal equilibrium}

The classical Smoluchowski limit is related to a
separation of time scales between fast equilibration of
momentum and slow equilibration of position. In this way,
the Fokker-Planck equation for the phase space
distribution can be adiabatically reduced to a
Smoluchowski equation for the marginal distribution in
position space \cite{risken}. For quantum dissipative
systems the expectation is that friction makes the system
to behave more classically so that for strong friction the
complicated path integral expression (\ref{sm4})
simplifies considerably. This is indeed the case as we
have shown recently \cite{ankerhold01}. In the sequel we
will briefly collect the main findings and, to keep things
as transparent as possible, consider a one-dimensional
system only moving in a sufficiently smooth potential
field $V(q)$.
\begin{figure}
\center
\includegraphics[width=8cm,draft=false]{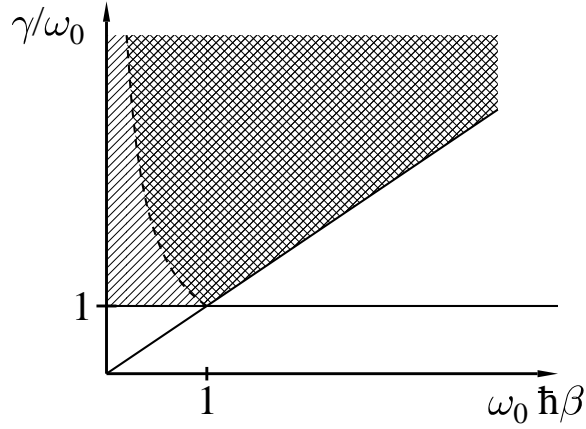}
 \caption{Quantum Smoluchowski range according to (\ref{sr2}) (shaded).
 The classical range $\gamma\hbar\beta< 1$ is simple shaded, the quantum range
 $\gamma\hbar\beta>1$ double shaded.}\label{fig1}
 \vspace*{0.0cm}
\end{figure}

A typical damping strength in the long time limit is defined as
\begin{equation}
\gamma\equiv\hat{\gamma}(0)=\lim\limits_{\omega\to
  0}\frac{I(\omega)}{M\omega}\label{sr1}
\end{equation}
where $\hat{\gamma}(\omega)$ is the Laplace transform of the classical
damping kernel $\gamma(s)$. For ohmic friction,
$I(\omega)=M\tilde{\gamma}\omega$, for instance, one finds
$\gamma=\tilde{\gamma}$. The same is true for the more realistic Drude
damping $I(\omega)=M\tilde{\gamma}\omega
\omega_c^2/(\omega^2+\omega_c^2)$ with cut-off frequency $\omega_c$. Now,
given a typical frequency $\omega_0$ of the bare system (e.g.\ its ground
state frequency) by strong damping we then mean (cf.~fig.~\ref{fig1})
\begin{equation}
\frac{\gamma}{\omega_0^2}\gg \frac{\hbar\beta}{2 \pi},
\frac{1}{\omega_c},\frac{1}{\gamma}\, .\label{sr2}
\end{equation}
Hence, we extend the time scale separation known from the
classical Smoluchowski range to incorporate the time scale
for quantum fluctuations $\hbar\beta$. As we will see,
this does not mean, however, that quantum effects are
negligible at all, since they are related to Matsubara
frequencies $\nu_n=2\pi n /\hbar\beta, n=1,2,3,\ldots$
which can be arbitrarily large.

The idea is now, to evaluate for strong friction
$\gamma/\omega_0\gg 1$ the path integral expression
(\ref{sm4}) on a coarse-grained time scale $s\gg
\hbar\beta, 1/\omega_c,1/\gamma$ and  $\tau\gg 1/\omega_c,
1/\gamma$. The consequences are the following: (i)
Non-diagonal elements of the reduced density matrix are
strongly suppressed during the time evolution, (ii) the
real time part of the kernel $K(s)$ becomes local on the
coarse grained time scale, and (iii) initial correlations
described by (\ref{sm3}) survive for times of order
$\gamma/\omega_0^2$  verifying that factorizing initial
states cannot be used.

In a first step, this program is applied to calculate the
thermal (unnormalized) equilibrium density matrix
\begin{equation}
\rho_\beta(\bar{q},\bar{q}')=\int {\cal D}[\bar{q}] \, {\rm
    e}^{i\bar{\Sigma}[\bar{q}]/\hbar}\label{sr3}
\end{equation}
with $\bar{\Sigma}[\bar{q}]=\Sigma[0,0,{\bar q}]$. Due to
(i) the strategy is to invoke a semiclassical type of
approximation by assuming self-consistently that
$|\bar{q}(\tau)-\bar{q}(0)|$ remains small on the time
interval $\hbar\beta$. In function space the path integral
is then dominated by the contributions of the minimal
action paths and Gaussian fluctuations around them. The
former ones are determined perturbatively up to corrections
of order $1/\gamma$ and
 its corresponding minimal action---conveniently
expressed in terms of the sum
$\bar{r}=(\bar{q}+\bar{q}')/2$ and difference coordinate
$\bar{x}=\bar{q}-\bar{q}'$--- follows as
\begin{equation}
\frac{-i}{\hbar}\bar{\Sigma}(\bar{r},\bar{x})=\beta
V(\bar{r})-\Lambda\beta^2\,
V'(\bar{r})^2+\frac{\Omega}{2\hbar^2}\,
\bar{x}^2+O\left(\frac{\Lambda}{\gamma}\right)
\label{minac}
\end{equation}
with
\begin{eqnarray}
\Lambda& =&\frac{2}{M\beta}\,
\sum_{n=1}^\infty\frac{1}{\nu_n^2+\nu_n\hat{\gamma}(\nu_n)}\ \ ,\nonumber\\
\Omega &=&\frac{M}{\beta}+\frac{2 M}{\beta}
\sum_{n=1}^\infty\frac{\hat{\gamma}(\nu_n)}{\nu_n+\hat{\gamma}(\nu_n)}\,
. \label{coeff}
\end{eqnarray}
Note that in $\Lambda$ the quadratic dependence on $\nu_n$
originating from the inertia part in the equation of
motion  and in $\Omega$ the cut-off frequency $\omega_c$
appearing in $\hat{\gamma}(\omega)$  are essential for the
convergence of the corresponding sums. This reveals a
fundamental difference to the classical Smoluchowski limit
and indicates the importance of quantum fluctuations even
for strong friction. From the above, one observes that
$\Lambda$ measures the typical strength of quantum
fluctuations in position and $\Omega$ is related (via the
identification $\hbar/\Delta x\to \Delta p$) with the
equilibrium variance in momentum $\Omega=\langle
p^2\rangle$.   Before we discuss further details, let us
arrive at the complete expression
 of the thermal equilibrium density matrix. The remaining Gaussian
fluctuations around the minimal action paths are treated
by switching to Matsubara frequency space.
Correspondingly, a straightforward calculation  provides
together with (\ref{minac})
\begin{equation}
\rho_\beta(\bar{x},\bar{r})= \frac{1}{Z} \, {\rm
e}^{-\beta V(\bar{r})-\Omega\, \bar{x}^2/2\hbar^2}\ {\rm
e}^{\Lambda\beta [ \beta V'(\bar{r})^2/2-3 V''(\bar{r})/2]}\, \label{equiden}
\end{equation}
where $Z$ denotes a proper normalization factor, e.g.\
$Z=\int dq \rho_\beta(0,q)$.  This is the first important
result of the strong friction analysis: The equilibrium
distribution of a strongly damped quantum system in
arbitrary (sufficiently smooth) potentials.  In
particular, for a harmonic oscillator the known result is
recovered \cite{weiss}. To be specific we consider Drude
damping $\hat{\gamma}(z)=\gamma\omega_c/(\omega_c+z)$ with
a high frequency cut-off $\omega_c\gg \gamma$. Then, the
functions $\Lambda$ and $\Omega$ can be expressed in terms
of $\Psi$ functions
\begin{eqnarray}
\Omega &\approx
&\frac{M\hbar\gamma}{\pi}\left[\Psi\left(\frac{\omega_c}{\nu}\right)-
  \Psi\left(\frac{\gamma}{\nu}+\frac{\gamma^2}{\nu\omega_c}\right)-
  \frac{\nu}{2\gamma}   +\frac{2\nu}{\omega_c}\right]\nonumber\\
\Lambda &\approx &\frac{\hbar}{M\gamma\pi}
\left[\Psi\left(\frac{\gamma}{\nu}\right)-C+\frac{\nu}{\gamma}\right]
 \label{sr4}
\end{eqnarray}
where $C=0.577\ldots$ is Euler's constant. In the high
temperature limit $\gamma\hbar\beta\ll 1$, one finds
$\Lambda\approx \hbar^2\beta/12 M$ and $\Omega\approx
M/\beta$ so that the Wigner transform of (\ref{equiden})
reduces to the classical phase space distribution. The
dependence on friction appears for lower temperatures as a
genuine quantum effect. Of particular interest is  the
limit $\gamma\hbar\beta\gg 1$ where we have
$\Lambda\approx (\hbar/M\gamma\pi) {\rm
log}(\gamma\hbar\beta/2\pi)$ and $\Omega\approx
(M\hbar\gamma/\pi){\rm log}(\omega_c/\gamma)$. Quantum
fluctuations in position are suppressed by friction, not
algebraically though, but much weaker. They also show a
nonlinear dependence on $\hbar$ meaning that for
$\gamma\hbar\beta\gg 1$ we work nevertheless in a deep
quantum domain. The same can be seen from the momentum
variance which grows with friction to guarantee
Heisenberg's uncertainty relation. Accordingly, as assumed
above, off-diagonal elements of the distribution
$\rho_\beta(\bar{q},\bar{q}')$ are strongly suppressed
with $|\bar{x}|$ being of order $1/\sqrt{\gamma {\rm
log}(\omega_c/\gamma)}$ or smaller. Concluding this
section, we mention that higher order corrections in
$\Lambda$ are associated with higher than second order
derivatives of the potential $V(q)$.

\subsection{Quantum Smoluchowski equation}

The analysis of the previous section already indicated the
strong suppression of off-diagonal elements of the density
distribution in thermal equilibrium. While in principle we
could now proceed to study the reduced dynamics in real
time for the total distribution $\rho(q_f,q_f',t)$
\cite{smolphase}, we concentrate in the sequel on the
quantum analog of the classical Smoluchowski limit and
thus restrict ourselves to the time evolution of its
diagonal part, the position distribution
$P(q_f,t)=\rho(q_f,q_f,t)$ \cite{ankerhold01}. In fact,
one can show that off-diagonal elements relax to thermal
equilibrium with respect to the instantaneous position of
the Brownian particle on  a time scale of order $1/\gamma$
\cite{annals}. For this purpose it is convenient to
introduce sum and difference paths also in real time
\begin{equation}
r(s)=[q(s)+q'(s)]/2\ ,\ \ \ \ x(s)=q(s)-q'(s)\label{dyn1}
\end{equation}
with corresponding fixed end-coordinates $r_f=r(t)=q_f,
x_f=x(t)=0$ and initial coordinates  $r_i=r(0), x_i=x(0)$
distributed according to the initial state (\ref{sm3}).
Already the classical Smoluchowski limit reduces the set
of acceptable initial states to those with momenta
sufficiently bounded from above. In the quantum domain,
the detailed analysis reveals that the $x_i$-dependence of
the preparation function $\lambda(x_i,r_i)$ must be
sufficiently smooth within the range
$|x_i/\sqrt{\hbar/M\gamma}|{\textstyle
 {\lower 2pt \hbox{$<$} \atop \raise 1pt \hbox{$ \sim$}}}
\sqrt{\hbar\beta/(\gamma/\omega_0^2)}$ such that we may put
 $\lambda(x_i,r_i)\approx \lambda(0,r_i)$.

Now, let us first look at the leading order contribution
where quantum fluctuations in position are neglected,
i.e.\ $\Lambda=0$. This is expected to lead us back to the
classical Smoluchowski equation, for $\gamma\hbar\beta\gg
1$ though, i.e.\ far from the classical limit
$\gamma\hbar\beta\ll 1$.  As above, we assume excursions
from diagonality, encoded in the $x$-paths, to be small,
roughly at most of order $1/\sqrt{\gamma}$. By expanding
the effective action $\Sigma[\bar{q},x,r]$ up to second
order in $x$ we arrive at a solvable Gaussian path
integral for the $x$-paths. We put $x(s)=x_i+\delta x(s)$
with deviations $\delta x(0)=\delta x(t)=0$ and obtain to
leading order
\begin{equation}
\int {\cal D}[x] \ {\rm
  e}^{i(\Sigma[\bar{q},x,r]-\Sigma[\bar{q},0,0])/\hbar}\approx \delta(x_i)\,
  {\rm e}^{-S[r]/4M\gamma k_{\rm B} T}\label{dyn2}
\end{equation}
with the action
\begin{equation}
S[r]=\int_0^t ds\, \left[M\gamma \dot{r}+V'(r)\right]^2\, \label{dyn3}.
\end{equation}
Thus, the imaginary time path contribution follows from the leading order
expression of (\ref{equiden}) for $\bar{x}=x_i=0, r_i=\bar{r}$. One
is left with the $r$-path integral which, written in form
of a propagator, appears as
\begin{equation}
P(q_f,t)=\int dq_i\, G(q_f,t,q_i)\, P(q_i,0)\label{dyn4}
\end{equation}
where in $G(\cdot)$ the paths $r(s)$ run from $r(0)=q_i$
to $r(t)=q_f$, each one weighted with the exponential on
the right hand side of (\ref{dyn2}). One this way ends up
with the path integral representation of the classical
Smoluchowski dynamics where $P(q,t)$  obeys
\begin{equation}
\frac{\partial P(q,t)}{\partial t}= \frac{1}{M\gamma}
\frac{\partial}{\partial q}\, L_{\rm cl}\, P(q,t)\label{dyn5}
\end{equation}
with the classical Smoluchowski operator $L_{\rm cl}=V'+k_{\rm B} T
\partial_q$. Indeed, from this derivation one finds $x(s)^2$ to be at
most of order $\hbar^2\beta/M\gamma t$.

To include quantum fluctuations to lowest order in
$\Lambda$, we could proceed to directly calculate
corrections to the propagator as in \cite{ankerhold01},
which is rather involved, however. Here, we follow a
somewhat simpler route and first exploit that the flux
vanishes in thermal equilibrium. Accordingly, one writes
 $\partial_t P=(1/\gamma M)\partial_q (1+\delta f)
\tilde{L}_{\rm qm}\, P$ with a dynamical correction $\delta
f$, and an operator $\tilde{L}_{\rm qm}$ determined
perturbatively up to order $\Lambda$ by $\tilde{L}_{\rm
qm}\rho_\beta=O(\Lambda^2)$ with the equilibrium density
specified in (\ref{equiden}). The latter one  already
contains the dominant part of the quantum fluctuations in
the QSE (see below). To verify that $\delta f$ can be
disregarded, it suffices to study the path integral
dynamics in the time range
$\hbar\beta,1/\gamma,1/\omega_c\ll t\ll
\gamma/\omega_0^2$  where it becomes effectively Markovian.
Then, along the lines described  above a semiclassical
approximation applies and leads to a propagator including
$\Lambda$ dependent corrections. From this, one finds
$\delta f\propto O(\Lambda/\gamma^2)$ and thus to be
negligible. As the result, the equation for
  $P(q,t)$ is of the form of
  the classical Smoluchowski equation (\ref{dyn5}) but with  $L_{\rm
    cl}$ replaced by $\tilde{L}_{\rm qm}$, i.e.,
\begin{equation}
\tilde{L}_{\rm qm}=U'_{\rm eff}(q)+k_{\rm B}T\partial_q
\tilde{D}(q)\label{dyn6}
\end{equation}
where we introduced an effective force field and a diffusion
coefficient, respectively,
\begin{equation}
U_{\rm eff}(q)=V'(q)+\frac{\Lambda}{2} V''(q)\ ,\ \ \ \
\tilde{D}(q)=1+\Lambda\beta V''(q)\, .\label{dyn7}
\end{equation}
The expressions (\ref{dyn6}) and (\ref{dyn7}) determine
the dynamics of overdamped quantum systems at lower
temperatures. The corresponding quantum Smoluchowski
equation has still one deficiency though, which is directly
related to the fact that it has been derived
perturbatively. Namely, it conserves thermodynamic
symmetries, as e.g.\ vanishing stationary currents in
thermal equilibrium, only to order $\Lambda$ meaning that
in a direct numerical integration higher order terms  may
give rise to unphysical results. The most obvious way to
see this, is to consider the stationary current $J_{\rm
st}=\lim_{t\to \infty}\langle \dot{q}\rangle/a$ in a
periodic, asymmetric potential $V(q+a)=V(q), V(q)\neq
V(-q)$ with {\em no} external bias (ratchet potential)
\cite{macch}.
 It is determined from the
stationary distribution $P_{\rm st}$ via $L_{\rm qm}P_{\rm st}=-\gamma M
J_{\rm
 st}$ as
\begin{equation}
J_{\rm st}=\frac{1-{\rm e}^{-\beta\tilde{\psi}(a)}}{\gamma
M\beta}\left[\int_0^a dq\,  \frac{{\rm
e}^{-\beta\tilde{\psi}(q)}}{\tilde{D}(q)} \, \int_q^{q+a} dy\, {\rm
e}^{\beta\tilde{\psi}(y)}\right]^{-1}.\label{qse3}
\end{equation}
where $\tilde{\psi}(q)=\int_0^q dy U_{\rm
eff}'(y)/\tilde{D}(y)$. Apparently,  the diffusion
coefficient in (\ref{dyn7}) leads to
$\tilde{\psi}(a)=O(\Lambda^2)$ and thus to $J_{\rm st}\neq
0$. The strategy to cure this problem is to look for a
"uniformization" of the operator $\tilde{L}_{\rm qm}$ by
finding a proper diffusion coefficient $D(q)$ which
coincides with $\tilde{D}$ in order $\Lambda$, leads to the
equilibrium distribution (\ref{equiden}), and, of course,
respects $J_{\rm st}=0$ in thermal equilibrium. The only
functional form consistent with these constraints brings us
to
\begin{equation}
D(q)=1/[1-\Lambda\beta V''(q)]\ .\label{dyn8}
\end{equation}
We note  that $D$ can be seen as a Pad{\'e} approximant to
a series in $|\Lambda\beta V''|<1$ starting with the terms
included in $\tilde{D}$. The corresponding expression
(\ref{qse3}) for the stationary current is now given by
the potential $\tilde{\psi}(q)=\psi(q)-\psi(0)$ with
\begin{equation}
\psi(q)=U_{\rm eff}(q)-\frac{\beta
  \Lambda}{2}\left[V'(q)^2+(\Lambda/2) V''(q)^2\right]\ , \label{dyn9}
\end{equation}
which obeys full periodicity in agreement with the correct symmetry
$J_{\rm st}=0$. Further, $P_\beta(q)\propto \exp[-\beta
  \psi(q)]/D(q)$
coincides with $\rho_\beta(0,q)$ (\ref{equiden}) up to corrections of
order $\Lambda^2$ in the exponential  which is consistent with our
perturbation theory.

Eventually, we arrive at the quantum generalization of the
classical Smoluchowski equation, the so-called quantum
Smoluchowski Equation (QSE),
\begin{equation}
\frac{\partial P(q,t)}{\partial t}=\frac{1}{\gamma
  M}\frac{\partial}{\partial q}\left[U_{\rm eff}(q)+k_{\rm B}
  T\frac{\partial}{\partial q}D(q)\right]\, P(q,t)\label{dyn10}
\end{equation}
with $U_{\rm eff}$ specified in (\ref{dyn7}) and $D$ in
(\ref{dyn8}). The quantum analog to the classical Langevin
equation in the strong damping limit  follows (in the Ito
sense) as
\begin{equation}
M\gamma \dot{q}+U_{\rm eff}'(q)={\xi(t)}{\sqrt{D(q)}}\label{dyn11}
\end{equation}
with Gaussian white noise $\langle \xi(t)\rangle=0, \langle
\xi(t)\xi(t')\rangle= 2 M\gamma k_{\rm B} T\delta(t-t')$.
Eqs. (\ref{dyn10}) and (\ref{dyn11}) are the main results of this study.
They describe the reduced dynamics of an overdamped quantum system
from high down to very low temperatures and
show that the corresponding quantum stochastic process
 is equivalent to a classical process in an effective potential
and with multiplicative noise.

\section{Applications of the Quantum Smoluchowski
Equation}\label{applications}

The QSE plays in the strong friction limit a similar role
as the quantum master equations in the weak damping limit,
since it allows for explicit results in strongly condensed
phase systems. Three examples from chemistry, mesoscopic
physics, and soft matter will be presented in this second
part.

\subsection{Escape from a Metastable Well}\label{escape}

The problem of escape from a metastable well can be found
in a variety of realizations including chemical reactions,
diffusion in solids, or nuclear fission processes, to name
but a few (see also the contribution by H{\"a}nggi and
Ingold \cite{gert}). Its particular feature is a
separation of time scales between local well motion and
long time decay characterized by a decay rate.
 Classically,
 a first thorough analysis has been performed in a seminal work by
 Kramers \cite{kramers} with substantial extensions since then \cite{hangg90}.
  For quantum
 mechanical systems a large amount of research has been done in the
 1980s in the context of macroscopic quantum tunneling and the field has
 gained renewed interest recently due to its importance for solid-state based
 implementations of qubits. Here, we look at the problem at strong
 friction and low temperatures.

The archetypical situation is the following: In a
metastable potential $V(q)$ a barrier with height $V_b\gg
k_{\rm B} T, \hbar\omega_0$ separates a well region (well
frequency $\omega_0$) from a continuum. Initially
particles stay in local thermal equilibrium  inside the
well. As time elapses, particles surmount the barrier and
for intermediate times (plateau range)  their position
distribution becomes quasi-stationary $P(q,t)\to P_{\rm
st}(q)$ describing a constant flux across the barrier
\begin{equation}
J_{\rm st}=-\frac{1}{\gamma M}L_{\rm qm}P_{\rm st}\,
.\label{exam1}
\end{equation}
 The decay rate follows from
\begin{equation}
\Gamma=\frac{J_{\rm st}}{Z_{\rm well}}\label{exam2}
\end{equation}
with the well population $Z_{\rm well}$. For strong friction the
changeover from quasi-equilibrium in the well around $q=0$ and
nonequilibrium on the other side of the barrier is restricted to the
vicinity of the barrier top located at $q_b>0$. Hence, the stationary
distribution takes the form $P_{\rm st}(q)=P_\beta(q)\, g_{\rm st}(q)$
with a form factor obeying $g_{\rm st}(q)\to 1$ in a close range to the
left of $q_b$ and $g_{\rm st}(q)\to 0$ in a close range to the right of
$q_b$. In fact, from (\ref{exam1}) a direct integration yields
\begin{equation}
P_{\rm st}(q)=\frac{M\gamma\beta J_{\rm st}}{D(q)}\ {\rm
e}^{-\beta\psi(q)}\, \int_q^\infty dy\ {\rm e}^{\beta\psi(y)}\label{exam3}
\end{equation}
with $\psi$ specified in (\ref{dyn9}). We note in passing
that this distribution is identical to that obtained from
real-time calculations \cite{hangg90}. Now, inserting
$Z_{\rm well}=\int_{-\infty}^{q_b} dq P_{\rm st}(q)$ into
(\ref{exam2}) and by invoking a harmonic approximation
around the well minimum and the barrier top according to
the above discussion, we obtain the rate in the quantum
Smoluchowski range as
\begin{equation}
\Gamma_{\rm QSR}=\frac{\sqrt{V''(0) |V''(q_b)|}}{2\pi M\gamma}\ {\rm
e}^{-\beta V_b}\ {\rm e}^{\beta \Lambda (V''(0)+|V''(q_b)|)}\, .
\label{exam4}
\end{equation}
\begin{figure}
\center
\includegraphics[width=8cm,draft=false]{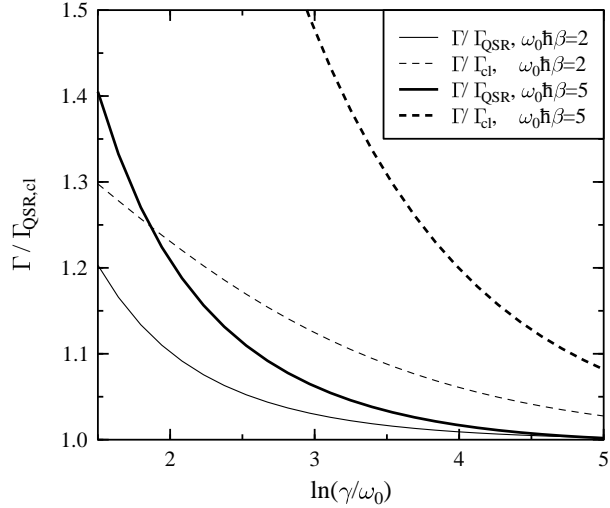}
 \caption{Ratio of the exact rate with the classical (dashed) and the QSE
 (solid)
 rate vs. friction for $|V''(q_b)|=V''(0)=M\omega_0^2$.}\label{fig2}
 \vspace*{0.0cm}
\end{figure}
In this expression the second exponential accounts for
quantum fluctuations, while the first factors coincide
with the overdamped Kramers rate. Note that $\Lambda$
dependent terms enter exponentially and thus substantially
enhance the quantum rate compared to the classical one
(see fig.~\ref{fig2}). Particularly in the high
temperature domain the rate enhancement is damping
independent and takes the form already derived by Wigner
\cite{weiss}. In the low temperature domain, however, a
complicated dependence on damping appears associated with
a strong increase of the rate compared to its classical
value. A quantum rate expression that becomes exact in the
semiclassical limit of a high barrier has been derived in
\cite{wolynes}. For strong friction this formula reduces to
(\ref{exam4}) and already for moderate friction agrees
well with the QSE-result, fig.~\ref{fig2}. Remarkably, the
rate enhancement is observable already at relatively high
temperatures $\sqrt{V''(0)/M}\hbar\beta<1$ provided
damping is sufficiently strong to guarantee
$\gamma\hbar\beta\gg 1$.

\subsection{Phase Diffusion and Charging Effects in Josephson
Junctions}\label{jj}

Since the discovery of the Josephson effect, devices based
on Josephson junctions (JJ) have revealed an extraordinary
wealth of phenomena studied theoretically and
experimentally as well \cite{barone,schon,nato}. Recent
realizations of the underlying system, two superconducting
domains separated by a tunnel barrier, comprise
superconducting atomic contacts \cite{atomic} and
solid-state based quantum bits \cite{qubit}. In essence,
two parameters determine the dynamics, namely, the
coupling energy of the adjacent domains (Josephson energy)
$E_{\rm J}$ and the charging energy of the contact $E_{\rm
c}=2 e^2/C$ of a junction with capacitance $C$. It turns
out that the competition between these two scales is
crucially influenced by the electromagnetic environment
surrounding the junction, i.e.\ its impedance which in the
simplest case is given by an ohmic resistor with
resistance $R$.

In the context considered here, a particular feature of a
JJ is the fact that its dynamics can be visualized as a
diffusive motion of a fictitious particle (RSJ model)
\cite{barone}. This mapping is due to the famous Josephson
relations
\begin{equation}
I_{\rm S}=I_{\rm c} \sin(\phi)\ \ ,\ \
\dot{\phi}(t)=\frac{2 e}{\hbar} V(t)\label{jj1}
\end{equation}
with $I_{\rm c}=(2 e/\hbar) E_{\rm J}$, the difference
between left and right superconducting phases $\phi$, and
where the phase velocity is related to the voltage drop
$V(t)$ across the junction. Charge transfer captured by
these relations corresponds to a phase coherent
Cooper-pair current. With the translation rules to the
mechanical analog
\begin{equation}
M=\left(\frac{\hbar}{2 e}\right)^2 C\ ,\ \
\gamma=\frac{1}{R C}\,  \label{jj3}
\end{equation}
the diffusive dynamics of the phase is determined by a
generalized Langevin equation. The case of very small
capacitance corresponds to very strong friction so that
for a current biased JJ the classical time evolution reads
\begin{equation}
M\gamma \dot{\phi}+d U(\phi)/d\phi=\xi(t)\, \label{jj2}
\end{equation}
with the Josephson potential $U(\phi)=-E_{\rm
J}\cos(\phi)-E_{\rm I}\phi$ where the energy $E_{\rm I}$ is
related to the bias current $(\hbar/2 e) I$, and current
noise $\langle\xi(t)\rangle=0,
\langle\xi(t)\xi(t')\rangle=(2\gamma/\beta)\delta(t-t')$.

This classical Smoluchowski dynamics has been studied in
detail already in the 60s \cite{ivanchenk,halperin}. The
QSE developed above, provides a generalization of this
description to lower temperatures where charging effects
enter the game \cite{josephsmolu}. In fact, it has been
shown in the context of single charge tunneling that for
sufficiently low temperatures these lead to Coulomb
blockade associated with incoherent transfer of Cooper
pairs through the junction \cite{udo}.

We start by translating the constraints  for the QSE dynamics to
the case of JJs. Here, in addition to the relation (\ref{sr2}) we
also have to take into account that typically the junction is
subject to an external voltage (or a corresponding current).
Accordingly, in order for the momentum $M \dot{\phi}$ to relax
within the $RC=1/\gamma-$time to a Boltzmann-like distribution
around $\langle \dot{\phi}\rangle$, the external voltage $V$ is
restricted by $e V\ll \hbar\gamma$. By combining
$\gamma^2/\omega_{\rm 0}^2\gg 1, \gamma\hbar\beta$, where
$\omega_{\rm 0}=\sqrt{2 E_{\rm c} E_{\rm J}}/\hbar$ is the plasma
  frequency of the unbiased JJ, with this latter condition and
  expressing them in
junction parameters we expect the QSE to capture quantum phase
diffusion in JJ if
\begin{equation}
\frac{E_{\rm c}}{E_{\rm J}2\pi^2\rho^2}\gg
  1,\ \frac{\beta E_{\rm c}}{2\pi^2\rho},\ \frac{V}{R I_{\rm c}}\,
  .\label{jj4}
\end{equation}
In the above, $\rho=R/R_Q$ with the resistance quantum
$R_Q=h/4 e^2$. Since typically $\rho\ll 1$ the above
condition allows for a broad range of values for $E_{\rm
c}/E_{\rm J}$,  $\beta E_{\rm J}$, and also large voltages
$V/R I_{\rm c}$. We note in passing that the above
relation also ensures that the actual non-ohmic impedance
seen by the junction can effectively be treated as ohmic.

Now, with the Josephson relations (\ref{jj1}) the effective
potential and the corresponding diffusion constant
entering the QSE (\ref{dyn10}) are found as
\begin{equation}
U_{\rm eff}(\phi)=-E_{\rm J}^\star\, \cos(\phi)-E_{\rm
  I}\phi\ ,\ \ \ \ \ E_{\rm J}^\star=E_{\rm J}
  \left(1-\frac{\Lambda}{2}\right)\label{jj5}
\end{equation}
and
\begin{equation}
 D(\phi)=[1-\theta\cos(\phi)]^{-1}\ ,\ \ \ \ \ \theta=\Lambda\beta
 E_{\rm J}\, .
\label{jj6}
\end{equation}
From the crucial $\Lambda$-function specified in (\ref{sr4}) and reading
here
\begin{equation}
\Lambda=2 \rho\, \left[c+\frac{2 \pi^2 \rho}{\beta E_{\rm
c}}+\Psi\left(\frac{\beta E_{\rm c}}{2 \pi^2\rho }\right)\right]\,
\label{jj6a}
\end{equation}
one sees immediately that $\beta E_{\rm
c}/\pi\rho=\gamma\hbar\beta$ controls the changeover from
classical to quantum Smoluchowski dynamics. In particular,
it turns out that  barrier related quantum fluctuations
incorporated in $E_{\rm J}^\star$ dominate in the limit of
$\beta E_{\rm J}\ll 1$, while diffusion related ones,
accounted for by the parameter $\theta$, prevail in the
opposite range $\beta E_{\rm J}\gg 1$.

The important observable is now the response of the JJ to an external bias
current, namely, the average voltage $\langle V\rangle$. It is related to
the steady state current via $\langle V\rangle=(\hbar/2\pi)\lim_{t\to
\infty}\langle\dot{\phi}\rangle=2\pi J_{\rm st}$. $J_{\rm st}$ reads as in
(\ref{qse3}) with periodicity $a=2\pi$, the diffusion constant from
(\ref{jj6}),  and the potential (\ref{dyn6}) adapted according to
(\ref{jj5}) and (\ref{jj6}). One then finds the current
voltage--characteristics of a current biased junction to read
\begin{equation}
\langle V\rangle = \frac{ \rho\, \pi}{\beta e}\frac{1-{\rm e}^{-2\pi
\beta E_{\rm I}}}{T_{\rm qm}}\ . \label{jj8}
\end{equation}
Here, the nominator $T_{\rm qm}$ results from normalizing
the steady state phase distribution to 1 and can be
written as
\begin{eqnarray}
T_{\rm qm}&=&\frac{1}{2\pi}\int_0^{2\pi}d\phi
\int_0^{2\pi} d\phi'\, {\rm e}^{-\beta E_{\rm I} \phi}\,
{\rm e}^{-2\beta E_{\rm
J}^\star\cos(\phi')\sin(\phi/2)}\nonumber\\
&&\times [1-\theta \sin(\phi'-\phi/2)]\,  {\rm e}^{2\beta\theta
\xi(\phi,\phi')}\label{jj9}
\end{eqnarray}
with
\begin{equation}
\xi(\phi,\phi')=  \sin(\phi')\sin(\phi/2)\left[E_{\rm I}+
E_{\rm J}^\star \cos(\phi') \cos(\phi/2)\right]\,
.\label{jj10}
\end{equation}
\begin{figure}
\center
\includegraphics[width=8cm,draft=false]{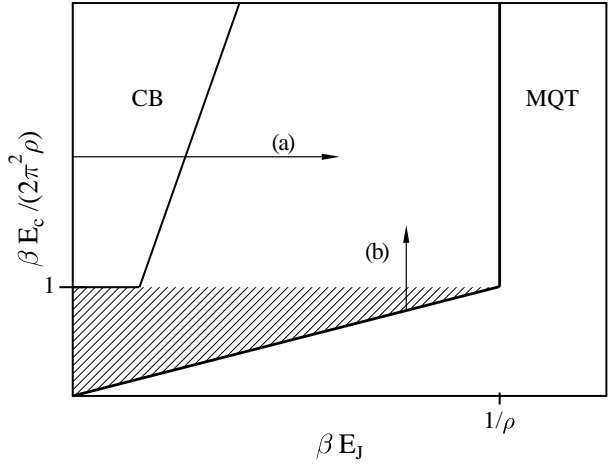}
 \caption{Range of the QSE  for a
 JJ with $\rho\ll 1$,  $V/R I_{\rm c}<1$. The classical IZT range
 (shaded)
 and the domains of Coulomb blockade (CB) and
 macroscopic quantum tunneling (MQT) are indicated. The QSE is applicable above the
 thick line, see (\ref{jj4}), and the arrows illustrate various
 changeovers discussed in the text.}\label{fig3}
 \vspace*{0.0cm}
\end{figure}

\begin{figure}
\center
\includegraphics[width=8cm,draft=false]{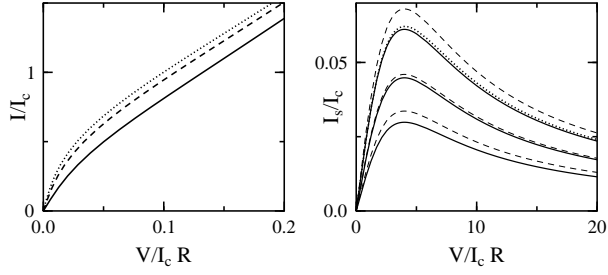}
  \caption{Left: Current-voltage characteristics for
 $\beta E_{\rm J}=2$, $\rho=0.04$. The quantum results [$\beta
E_{\rm c}=1$ (dashed), $\beta E_{\rm c}=20$ (solid)] are
shown with the classical one ($\beta E_{\rm c}=0$,
dotted). Right: Supercurrent vs. voltage for $\beta E_{\rm
J}=0.25$, $\rho=0.04$.  The classical result (dotted, from
\cite{ivanchenk}) and the CB expression [dashed, from
\cite{gli}] are depicted together with the QSE result
[solid, from \cite{josephsmolu}] for $\beta E_{\rm c}=0.15$
(top), $\beta E_{\rm c}=20$ (middle), $\beta E_{\rm
c}=1000$ (bottom).} \label{fig4}
 \vspace*{-0.0cm}
\end{figure}

The expression (\ref{jj8}) together with (\ref{jj9}) is
the central result from which various known findings can
be derived as limiting cases (see fig.~\ref{fig3}). (i)
For $\beta E_{\rm c}/\rho\ll 1$ the function $T_{\rm qm}$
reduces to its classical form ($E_{\rm J}^\star\to E_{\rm
J}, \theta\to 0$) and the classical Ivanchenko-Zil'berman
Theory (IZT) \cite{ivanchenk} is recovered. (ii) In the
low temperature domain $\beta E_{\rm c}/\rho\gg 1$, but
for smaller couplings $\beta E_{\rm J}< 1$, we have
$\theta\ll 1$ so that diffusion related quantum
fluctuations are negligible. As a consequence, quantum
effects in the transport can be described by a renormalized
coupling energy $E_{\rm J}^\star$. This important
extension of IZT has first been derived in \cite{gli}
based on a direct evaluation of the real-time path
integral expression. The supercurrent across the junction
coincides with results from Coulomb blockade (CB) theory
(cf.\ fig.~\ref{fig4}), thus describing an incoherent
transfer of charges. (iii) For $\beta E_{\rm c}/\rho\gg 1$
and sufficiently larger couplings $\beta E_{\rm J}>1$,
coherent Cooper pair tunneling exists.
Then, for $\alpha=I/I_{\rm c}<1$ occasional phase slips
occur and lead to the voltage
\begin{equation}
\frac{\langle V\rangle}{R I_{\rm
c}}=\frac{\sqrt{1-\alpha^2}}{2\pi}\ {\rm
    e}^{ -2\beta E_{\rm J}
    (1-\alpha^2)^{3/2}/(3\alpha^2)}\ {\rm e}^{ 2\theta
    \sqrt{1-\alpha^2}}\, \label{jj11}
\end{equation}
 which via $\theta$ is
affected by diffusion related quantum fluctuations. As can
be observed in fig.~\ref{fig3}, the result (\ref{jj11})
tends for $\theta\to 0$ to classical thermal activation
over the barriers of the washboard potential $U(\phi)$,
where quantum corrections are of the known damping
independent form \cite{feynman2}. At lower temperatures,
i.e.\ for finite $\theta$, they show a complicated
dependence on $\rho$ and capture the precursors of
macroscopic quantum tunneling (MQT) found at very low
temperatures \cite{kur}. Thus, the central result
(\ref{jj8}) fills the gap between established results in
different transport domains: On the one hand,  for fixed
$\beta E_{\rm c}/\rho>1$ it leads with increasing $\beta
E_{\rm J}$ from Coulomb blockade to coherent Cooper pair
tunneling [fig.~\ref{fig3}, arrow (a)]. On the other hand,
 for fixed $\beta E_{\rm J}>1$ it connects with varying
 $\beta E_{\rm c}/\rho$
 classical thermal activation with MQT [fig.~\ref{fig3}, arrow (b)]. Apart
 from limiting cases, (\ref{jj8}) is easily evaluated
 numerically; some results
 are shown in fig.~\ref{fig4}. Hence, the QSE approach allows to
give, in a seemingly transparent manner, a complete
description throughout a broad range in parameter space
and must be supplemented only for very low temperatures
($T\to 0$) by more sophisticated techniques \cite{gliprl}.

\subsection{Electron transfer in  condensed phase}\label{et}

\begin{figure}
\center
\includegraphics[width=7.5cm,height=6cm,draft=false]{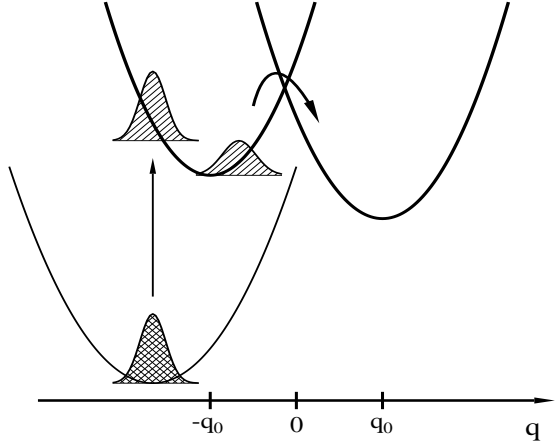}
 \caption{Diabatic potential surfaces for the reaction
  coordinate (RC) in a ET process. A wave packet is excited by a laser
  pulse from a dark state (thin line) to the donor state (thick line,
  minimum at $-q_0$). There it evolves on the coupled donor and acceptor
  (thick line,
  minimum at $+q_0$) surfaces where the ET occurs in the
  Landau-Zener region near $q=0$.} \label{fig5}
\end{figure}

The most prominent example of molecular electron transfer
(ET) is the primary step of photosynthesis, but it is of
fundamental relevance in a large variety of chemical and
biological compounds \cite{marcus,jortner}. In the
simplest case, two electronic sites denoted as donor and
acceptor, respectively, interact with a bath of harmonic
oscillators (spin-boson model \cite{leggett87}).
Completely equivalent is a formulation where the
electronic system is coupled to a collective vibronic
degree of freedom (Reaction Coordinate, RC) embedded in a
thermal environment \cite{garg}, see fig.~\ref{fig5}. The
theoretical foundations to describe these systems have
already been laid in the 1980s with the pioneering work by
Marcus et al.\ \cite{marcus}. It turned out that
qualitatively two dynamical domains must be distinguished:
For vibronic dynamics fast compared to the bare ET,
characteristic for long-distance ET, one speaks of
nonadiabatic reactions; in the opposite case of very slow
vibronic motion, realized e.g.\ in mixed-valence
compounds, the reaction is said to be adiabatic.
Accordingly, for a donor-acceptor system transfer rates
have been successfully calculated in the nonadiabatic
regime by invoking golden rule techniques and for
adiabatic reactions  at sufficiently high temperatures by
Kramers' flux over population method \cite{weiss}. These
quite different approaches reflect the different physical
processes that control the transfer, namely, in the former
case the electronic coupling between diabatic
donor/acceptor states,  in the latter  one the sluggish
activated bath motion on the lower adiabatic surface. For
sufficiently high temperatures and  slow to moderately slow
bath modes, first Zusman \cite{zusman} and later Garg
\cite{garg} derived equations of motion for the electronic
dynamics interacting with a damped RC treated in the
classical Smoluchowski limit. Since these Zusman Equations
(ZE) allow to derive a rate expression which to some extent
comprises adiabatic as well as nonadiabatic effects, it is
important to look for a low temperature extension
\cite{etsmolu}.

The corresponding Hamiltonian is of the form (\ref{sm1}) with a system
part $H_S=H_{\rm EL}+H_{\rm RC}$ where
\begin{equation}
H_{EL}
     =  -\frac{\hbar\epsilon}{2} \sigma_z
        -  \frac{\hbar\Delta}{2} \sigma_x \label{et1}
\end{equation}
represents the bare electronic two-state system (EL) as an artificial
spin-$\frac{1}{2}$ system with donor $|-\rangle$ and acceptor
$|+\rangle$,  a bias $\hbar \epsilon$, and an electron transfer coupling
$\hbar\Delta$; the second part
\begin{equation}
H_{RC}
     =      \frac{p^2}{2M}
        + \frac{M\omega_0^2}{2}q^2 -c_0 q \sigma_z\label{et2}
\end{equation}
describes a harmonic collective vibronic degree of freedom linearly
coupled to the electronic system and to a heat bath (see
fig.~\ref{fig5}). What one is interested in is the dynamics of a state
initially prepared on the donor surface. It is given by the exact path
integral expression for the reduced density matrix of the EL+RC system
$\rho_{\alpha \beta}(q,q',t)$, $\alpha,\beta\in\{-,+\}$ \cite{lucke}.

Now,  for strong damping (slow RC) a simplification
according to the scenario discussed above is applicable.
The difficulty then lies in the coupling between
overdamped RC and electronic motion  which basically makes
a transition from one electronic state to the other only
possible in the Landau-Zener (LZ) region where the
harmonic surfaces of the RC intersect (see
fig.~\ref{fig5}). Thus, three additional time scales enter
the game: The typical time for thermal activation into the
LZ range, $t_{\rm act}$, the typical time the RC diffuses,
after activation, in the LZ range, $t_{\rm LZ}$, and the
typical time the density matrix resides in a non-diagonal
state, $t_{\rm blip}$. It turns out that (\ref{sr2}) must
be extended to
\begin{equation}
\frac{1}{\gamma},\frac{\hbar\beta}{2\pi}\ll t_{\rm blip}\ll t_{\rm LZ}<
  \frac{\gamma}{\omega_0^2}\ll t_{\rm
  act}\, \label{et3}
\end{equation}
in order to reduce the path integral expression. This way,
one derives for the probabilities
$P_{\alpha\beta}(q,t)=\rho_{\alpha \beta}(q,q,t)$ the
extension of the ZE to lower temperatures, the Generalized
Zusman Equations (GZE),
\begin{eqnarray}
\dot{P}_{\mp\mp}(q,t)
    & = &   {\cal L}^\mp\ P_{\mp\mp}(q,t)
        \pm \frac{i\Delta}{2}\, \left[ P_{-+}(q,t)-P_{+-}(q,t) \right]
        \nonumber\\
\dot{P}_{\mp\pm}(q,t)
    & = &   {\cal L}^0 \ \, P_{\mp\pm}(q,t)
        \pm \frac{i\Delta}{2}\left[P_{--}(q,t)-P_{++}(q,t)
        \right]\nonumber\\
       && \mp i\, \left(\epsilon+\frac{2 c_0}{\hbar\kappa}\
          q \right)\ P_{\mp\pm}(q,t)
\, .\label{et3a}
\end{eqnarray}
Here,  the quantum Smoluchowski operators [see
(\ref{dyn10})] read
\begin{equation}
{\cal L}^{\eta}=  \frac{1}{M\gamma} \frac{\partial}{\partial q}\,
        \left[ M \omega_0^2 (q
        -\eta\, q_0) + \frac{\kappa}{\beta} \,
        \frac{\partial}{\partial q}\right]\label{et4}
\end{equation}
and describe the overdamped quantum dynamics of the
harmonic RC on the donor ($\eta=-$), the acceptor
($\eta=+$), and the averaged potential surface
($\eta=0$).  Note that the effect of quantum fluctuations
also shows up in the $c_0/\kappa$ dependent coupling terms
of the non-diagonal matrix elements. The coefficient
$\kappa$ contains the equilibrium variance of a damped
harmonic oscillator
\begin{equation}
\kappa=\frac{\langle q^2 \rangle}{\langle q^2
  \rangle_{cl}}\approx 1+M\omega_0^2 \beta \Lambda\, .\label{et5}
\end{equation}

\begin{figure}
\center \vspace*{0.5cm}
\includegraphics[width=8cm,draft=false]{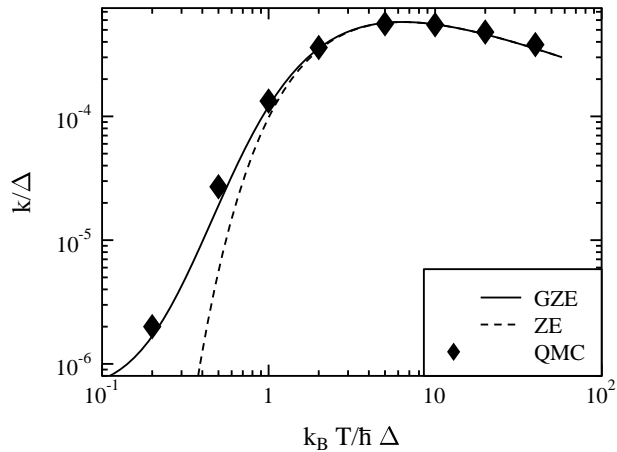}
 \caption{Electron transfer rates vs. temperature for a
symmetric system
  ($\epsilon=0$) according to the expression (\ref{et6}) (solid
  line), the classical result ($\kappa=1$, dashed line),  and
  precise Quantum Monte Carlo data (diamonds, from
  \cite{lothar}). $E_{\rm r}/{\hbar\Delta}=10$
  and $\Delta/(\omega_0^2/\gamma)=2$.} \label{fig6}
\end{figure}

By invoking again the time scale separation (\ref{et3}) the
electron transfer rate can be derived from the GZE
\cite{etsmolu,cukier,goychuk}. In case of a vanishing bias
one finds for the total (forward+backward) rate
\begin{equation}
k=\frac{\Delta^2}{1+ g}\ \sqrt{\frac{\hbar^2 \pi \beta}{4\,E_{\rm
      r}/\kappa}}\
\exp\left(-\beta E_{\rm r}/4\kappa\right)\, \label{et6}
\end{equation}
with an adiabaticity parameter
\begin{equation}
g=\pi\,\kappa\ \frac{\Delta}{\omega_0^2/\gamma}\
\frac{\hbar\Delta}{E_{\rm r}}\, \label{et7}
\end{equation}
and with the reorganization energy $E_{\rm r}=2 c_0^2/M \omega_0^2$. For
$g\gg 1 $ one recovers the adiabatic, for $g\ll 1 $ the nonadiabatic rate
constant. The above rate expression looks like the classical
Marcus/Zusman result with a renormalized reorganization energy $E_{\rm
r}\to E_{\rm
  r}/\kappa$. Note that this
simple renormalization only appears if quantum fluctuations in the
$c_0$-coupling terms of the GZE are properly taken into account. Since
$\kappa\geq 1$ quantum fluctuations always {\em reduce} the effective
energy barrier which is to be surmounted. Of course, in the high
temperature limit ($\kappa\to 1$) we regain the known result, for lower
temperature, however,  significant deviations are observed.
 (i) The ratio $\kappa$ grows with decreasing temperature, at very low
temperatures roughly linearly with $\beta$. As a
consequence, keeping all other parameters fixed, $g$
becomes larger with lower $T$ meaning that one approaches
the range where the transfer is dominated by adiabatic
processes at smaller values of $\Delta$ compared to the
classical range. (ii) At lower $T$, the exponent in the
rate expression $\beta E_{\rm r}/\kappa$ tends to become
temperature independent in contrast to the classical
result. Since $\kappa$ enters the exponent in (\ref{et6}),
even relatively small deviations from the classical
behavior $\kappa/\beta=1/\beta$ substantially influence
the rate.
 In fact, if  formally the limit of very
low temperatures $\gamma\hbar\beta\gg 1, \omega_0\hbar\beta\gg
\gamma/\omega_0$ is taken [which in a strict sense
   is out of the range of validity of
 (\ref{et6})], $\kappa/\beta$ saturates and one obtains
\begin{equation}
\frac{\beta E_{\rm r}}{4\kappa} \to  \frac{\pi\gamma}{8\omega_0\, {\rm
  ln}(\gamma/\omega_0)}\,
  \frac{ E_{\rm
  r}}{\hbar\omega_0}\label{kap}
\end{equation}
which is identical to
  the overlap of two Gaussian wave packets
 with overdamped harmonic variance
 localized around $\mp c_0/M\omega_0^2$.
Hence, nuclear tunneling is included in the above rate
formula, at least in a nondynamical way. In comparison
with precise path integral Monte Carlo
  results (from \cite{lothar}) a remarkable agreement is seen
  over a broad temperature range, fig.~\ref{fig6}. At low $T$,
  the rate enhancement due to nuclear tunneling is substantial.

\section{Conclusions}

The description of quantum Brownian motion is still a
challenging problem. In general it does not allow for
"simple" solutions in terms of tractable time evolution
equations and even a direct numerical evaluation of the
exact path integral expression for the reduced dynamics is
not always feasible. While in the limit of weak friction
various master equations have successfully been applied in
the past, the opposite limit of strong friction has been
left basically untouched. However, the naive expectation is
that a strongly condensed phase environment quenches
quantum fluctuations, thus forcing the system to behave
more classically. In fact, this is only one facet of the
problem since quantum mechanically small fluctuations in
position are immediately accompanied by large fluctuations
in the conjugate momentum. A systematic simplification
within the path integral formulation leads indeed to an
equation of motion which to leading order coincides with
the classical Smoluchowski equation, but contains
substantial quantum corrections.

Applications for chemical, mesoscopic, and soft matter
systems have shown the decisive impact of corresponding
quantum mechanical fluctuations on transport properties
even up to high temperatures where thermal energies exceed
relevant time scales of the damped dynamics. So far, this
analysis has focused on one-dimensional systems, but by
exploiting the developed techniques extensions to two and
three dimensions are certainly feasible.

\acknowledgements

We are grateful for the financial support of the DAAD,
NSF, and the DFG. J.A.\ acknowledges a Heisenberg
fellowship of the DFG.

\end{document}